\documentclass[prl,twocolumn,superscriptaddress,showpacs,preprintnumbers]{revtex4}
\usepackage{amsmath,amssymb,graphicx,makeidx,dcolumn,epsfig}

\def\g{\gamma}

\def\o{\omega}

\def\D{\Delta}

\def\L{\Lambda}

\def\hs{\hspace}
\def\ol{\overline}
\def\no{\nonumber}
\def\lf{\left}
\def\rg{\right}

\begin{document}

\preprint{ADP-05-06/T617, JLAB-THY-05-316}
                                                                                                                             
\title{Spin-dependent structure functions in nuclear matter and the polarized EMC effect}

\author{I.C.~Clo\"et}
\email{icloet@physics.adelaide.edu.au}
\affiliation{Special Research Centre for the Subatomic Structure of
             Matter and      \\
             Department of Physics and Mathematical Physics, University of Adelaide,
             SA 5005, Australia}
\affiliation{Jefferson Lab, 12000 Jefferson Avenue, Newport News, VA 23606, U.S.A.}
\author{W.~Bentz}
\email{bentz@keyaki.cc.u-tokai.ac.jp}
\affiliation{Department of Physics, School of Science, Tokai University, 
                         Hiratsuka-shi, Kanagawa 259-1292, Japan}
\author{A.W.~Thomas}
\email{awthomas@jlab.org}
\affiliation{Jefferson Lab, 12000 Jefferson Avenue, Newport News, VA 23606, U.S.A.}

\begin{abstract}
An excellent description of both spin-independent and spin-dependent
quark distributions and structure functions has been obtained  
with a modified Nambu$-$Jona-Lasinio model,
which is free of unphysical thresholds for nucleon
decay into quarks $-$ hence incorporating an important aspect of confinement.
We utilize this model to investigate nuclear medium
modifications to structure functions and find that we are readily able 
to reproduce both nuclear matter saturation and the experimental $F_{2N}^A/F_{2N}$ ratio, 
that is, the EMC effect. 
Applying this framework to determine $g_{1p}^A$, we find that the ratio
$g_{1p}^A/g_{1p}$ differs significantly from 1, with the quenching caused
by the nuclear medium being about twice that of the spin-independent case.
This represents an exciting result, which if confirmed experimentally,
will reveal much about the quark structure of nuclear matter.
\vspace{1pc}
\end{abstract}


\maketitle

The discovery in the early 80's by the European Muon Collaboration (EMC) that 
nuclear structure functions differ substantially from those of free 
nucleons~\cite{Aubert:1983xm,Bodek:1983ec,Arnold:1983mw} 
caused a shock in the nuclear community.  Despite many 
attempts to understand this effect in terms of binding corrections 
it has become clear that one cannot understand it without a 
change in the structure of the nucleon-like quark clusters in 
matter \cite{Geesaman:1995yd,Smith:2002ci,Benesh:2003fk}.  
Mean-field models of nuclear structure built at the quark level, 
which have been developed over the past 15 years, are yielding a 
quantitative description of the EMC effect.
Most recently it has been demonstrated that at least one of these models
leads naturally to a Skyrme-type force, with parameters in agreement
with those found phenomenologically to describe a vast amount of
nuclear data~\cite{Guichon:2004xg}.

A second major discovery by the EMC concerned the 
so-called ``spin crisis''\cite{Ashman:1987hv}, which corresponds to the discovery that 
the fraction of the spin of the proton carried by its quarks is 
unexpectedly small.  This has led to major new insights into the famous  
$U$(1) axial anomaly, prompting many new experiments.  With this background,
it is astonishing that, in the 17 years since the discovery of the spin crisis,
there has been no experimental investigation of the spin-dependent 
structure functions of atomic nuclei.  Of course, such experiments are 
more difficult because the nuclear spin is usually carried by just 
a single nucleon and hence the spin dependence is an ${\cal O}(1/A)$ effect.
Nevertheless, as we shall see, such measurements promise another major 
surprise, with at least one model $-$ which reproduces the EMC effect 
in nuclear matter $-$ 
suggesting a modification of the spin structure function
of a bound proton in nuclear matter roughly twice as large as 
the change the spin-independent structure function.

Models of nuclear structure like the quark meson coupling (QMC) model, 
achieve saturation through the self-consistent change in the quark 
structure of the colorless, nucleon-like constituents $-$ in particular, 
through its scalar polarizability \cite{Guichon:2004xg,Chanfray:2004ev}. 
Physically the idea is extremely simple, 
light quarks respond rapidly to oppose an applied scalar field.  
Specifically, the lower components of the valence quark wave functions 
are enhanced and this in turn reduces the effective $\sigma N$ coupling.  
The fact that changes in the structure of bound nucleons are so 
difficult to find appears to be a result of this mechanism being extremely 
efficient and hence yielding only a small change in the dominant 
upper components of the valence quark wave functions.

On the other hand, the spin structure functions are particularly 
sensitive to the lower components and this is why the measurement 
of the spin-dependent EMC effect is so promising.
Our calculations are made within the framework developed by Bentz, 
Thomas and collaborators \cite{Bentz:2001vc,Mineo:2003vc},
in which proper-time regularization \cite{Schwinger:1951nm,Ebert:1996vx,Hellstern:1997nv}
is applied to the NJL model in order to simulate the effects of confinement.  
This model exhibits similar properties to the QMC model with the advantage 
that it is covariant. Once we include both scalar and 
axial-vector diquarks, it readily describes nuclear saturation at the correct 
energy and density.  Moreover it yields PDFs for the free nucleon \cite{cloet:2005}
which are in excellent agreement with existing experimental data.

We write the spin-dependent light-cone quark distribution of a nucleus with mass number $A$
and helicity $H$ as the convolution
\begin{align}
&\Delta f_{q/A}^{(H)}\left(x_A\right) = \nonumber \\
&\int\!\! dy_A\!\! \int\!\! dx\,\, \delta\!\left(x_A-y_A\, x\right) 
\Delta f_{q/N}\!\left(x\right)\, \Delta f_{N/A}^{(H)}\!\left(y_A\right),
\label{eqn:con}
\end{align}
where $\Delta f_{q/N}(x)$ is the spin-dependent quark light-cone momentum distribution 
in the bound nucleon, $\Delta f_{N/A}^{(H)}\!\lf(y_A\rg)$
the light-cone momentum distribution of the nucleon in the nucleus and $x_A \in \lf[0,A\rg]$ is 
the Bjorken scaling variable for the nucleus.
There have been numerous investigations of $\Delta f_{N/A}^{(H)}\lf(y_A\rg)$ 
\cite{Saito:2001gv} and it is straightforward to calculate for any particular 
nucleus. Examples of greatest experimental interest
would be single proton particle or hole states like $^7\textsl{Li}$, 
$^{11}\textsl{B}$ and $^{15}\textsl{N}$. 
In this analysis, as our primary focus is the change in $\Delta f_{q/N}$ in-medium,
we incorporate the Fermi motion effects on the bound proton structure function
by replacing $\Delta f_{N/A}^{(H)}(y_A)$ with the spin-independent distribution $f_{N/A}(y_A)$, calculated in 
infinite nuclear matter \cite{Mineo:2003vc}.
 
To calculate $\Delta f_{q/N}(x)$ in our model, it is convenient to express it in the form 
\cite{Jaffe:1985je,Barone:2001sp}
\begin{align}
\label{eqn:quarkstructure}
\Delta f_{q/N}(x) &= -i \int\frac{d^4k}{(2\pi)^4} \nonumber \\ 
&\hspace{16mm}\delta\!\left(z-\frac{k_-}{p_-}\right) \text{Tr}\left(\gamma^+\gamma_5\,M(p,k)\right), 
\end{align}
where $M(p,k)$ is the quark two-point function in the nucleon. Within any model that describes the nucleon as a bound 
state of quarks, this distribution 
function can be associated with a straightforward Feynman diagram calculation, 
where the propagators include the self consistent scalar and vector fields in the nucleus.

It is demonstrated in Ref.~\cite{Mineo:2003vc} that the in-medium changes to a
free nucleon quark distribution can be included as follows.
The effect of the scalar field is incorporated by simply replacing the free masses
with the effective masses in nuclear medium, giving the distribution
$\Delta f_{q/N0}(x)$ \footnote{The subscript 0 denotes the absence of any mean vector field.} and
the Fermi motion of the nucleon is included by convoluting this distribution ($\Delta f_{q/N0}(x)$) with the 
Fermi smearing function, $f_{N/A0}(\tilde{y}_A)$, producing the distribution
\begin{multline}
\Delta f_{q/A0}(\tilde{x}_A) = \int d\tilde{y}_A \int dz  \\ 
\delta\left(\tilde{x}_A - \tilde{y}_A\, z\right)\, \Delta f_{q/N0}(z)\, f_{N/A0}(\tilde{y}_A).
\label{eqn:convolution}
\end{multline}
The effect of the vector field is then incorporated via the scale transformation
\begin{multline}
\Delta f_{q/A}\left(x_A\right) = \frac{\varepsilon_F}{E_F}\, 
\Delta f_{q/A0}\left(\tilde{x}_A = \frac{\varepsilon_F}{E_F}x_A - \frac{V_0}{E_F}\right),
\label{eqn:vector}
\end{multline}
where $\varepsilon_F = \sqrt{p_F^2 + M_N^2} + 3V_0 \equiv E_F + 3V_0$ is the Fermi energy of the
nucleon, $p_F$ the Fermi momentum 
and $V_0$ is the zeroth component of the vector field felt by a quark.

%------------------------------------------------------------------------------------------
\begin{figure}[tbp] 
\centering\includegraphics[height=2.4cm,angle=0]{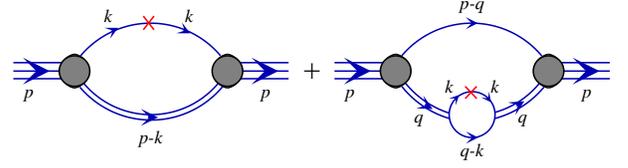}
\caption{Feynman diagrams representing the spin-dependent quark distributions in the nucleon, 
needed to determine $\Delta f_{q/N}(x)$, given in Eq.~(\ref{eqn:quarkstructure}). The single line
represents the quark propagator and the double line the diquark $t$-matrix. The shaded oval
denotes the quark-diquark vertex function and the operator insertion has the form
$\gamma^+\gamma_5\,\delta\!\left(x - \frac{k_-}{p_-}\right)\frac{1}{2}\left(1 \pm \tau_z\right)$.
The second diagram, which we refer to as the ``diquark diagram'',  
symbolically represents  two diagrams, 
each with the operator insertion on a different quark line within the diquark.}
\label{fig:feydiagrams}
\end{figure}
%-------------------------------------------------------------------------------------------

To calculate the spin-dependent quark distribution in 
the nucleon, $\Delta f_{q/N0}(x)$, we use the NJL model to describe the nucleon
as a quark-diquark bound state, taking into account both scalar $\lf(J^{\pi} = 0^+, T = 0, \text{colour}\,\ol{3}\,\rg)$
and axial-vector $\lf(J^{\pi} = 1^+, T = 1, \text{colour}\,\ol{3}\,\rg)$ diquark channels. 
Details of these free space calculations,
along with a description of the proper-time regularization scheme
used throughout this paper, may be found in Ref.~\cite{cloet:2005}.
In short, the quark distribution functions are determined from the 
Feynman diagrams of Fig.~\ref{fig:feydiagrams}, with the resulting
distribution $\Delta f_{q/N0}(x)$, having no support for negative
$x$. Hence, this is essentially a valence quark picture.  

By calculating these Feynman diagrams using the effective (density dependent) masses obtained from
the nuclear matter equation of state (discussed below) and performing the 
transformation, Eq.~(\ref{eqn:vector}),
to include the mean vector field, we obtain the spin-dependent $u$ and $d$ distributions
in a bound proton. Separating the isospin factors, gives 
\begin{align}
\label{eqn:delu}
\D u^A_v(x) &= \D f^s_{q/N}(x) + \frac{1}{2}\,\D f^s_{q(D)/N}(x) + \frac{1}{3}\, \D f^a_{q/N}(x)  \no \\
&\hs{6mm} + \frac{5}{6}\,\D f^a_{q(D)/N}(x) + \frac{1}{2\sqrt{3}} \D f_{q(D)/N}^m (x),  \\ 
\label{eqn:deld}
\D d^A_v(x) &= \frac{1}{2}\,\D f^s_{q(D)/N}(x) + \frac{2}{3}\,\D f^a_{q/N}(x)   \no \\
&\hs{6mm} + \frac{1}{6}\,\D f^a_{q(D)/N}(x) - \frac{1}{2\sqrt{3}} \D f_{q(D)/N}^m (x).
\end{align}
The superscripts $s$, $a$ and $m$ refer to the scalar, axial-vector and mixing
terms, respectively, the subscript $q/N$ implies a quark diagram and $q(D)/N$ a diquark diagram.
Because the scalar diquark has spin zero, we have $\D f^s_{q(D)/N}(x)=0$ and hence the
polarization of the $d$ quark arises exclusively from the axial-vector and the mixing terms. 

The NJL model is a chiral effective quark theory that is characterized by
a 4-Fermi contact interaction. Using Fierz transformations any 4-Fermi
interaction can be decomposed into various interacting $qq$ and $q\bar{q}$ channels \cite{Ishii:1995bu}.
The terms relevant to this discussion are
\begin{align}
{\cal L} &= \overline{\psi}\left(i\!\not\!\partial - m\right)\psi\, \no \\
&\hs{1mm}+\,G_\pi\left(\left(\overline{\psi}\psi\right)^2 - \left(\overline{\psi}\gamma_5\,\vec{\tau}\psi\right)^2\right)  
-\,G_\omega\left(\overline{\psi}\gamma^\mu\psi\right)^2 + \ldots \no \\
&\hs{1mm}+\,G_s \lf(\ol{\psi}\,\g_5 C \tau_2 \beta^A\, \ol{\psi}^T\rg)
                               \lf(\psi^T\,C^{-1}\g_5 \tau_2 \beta^A\, \psi\rg) \no \\
&\hs{1mm}+\,G_a \lf(\ol{\psi}\,\g_\mu C \tau_i\tau_2 \beta^A\, \ol{\psi}^T\rg) 
                               \lf(\psi^T\,C^{-1}\g^{\mu} \tau_2\tau_i \beta^A\, \psi\rg),
\label{eqn:lag}
\end{align}
where $m$ is the current quark mass, $\beta^A = \sqrt{\tfrac{3}{2}}\,\lambda^A~(A=2,5,7)$ are the 
colour $\ol{3}$ matrices and $C = i\g_2\g_0$.
In the $q \overline{q}$ channel
we include scalar, pseudoscalar and vector components and in the 
$qq$ channel we have the scalar and axial-vector diquarks. 
The scalar $q \overline{q}$ interaction term generates
the scalar field, that is, the constituent quark mass $M$ (vacuum value $M_0$) via the gap equation.
The vector $q \overline{q}$ interaction
will be used to generate the vector field in-medium. The $qq$ interaction terms give the diquark
$t$-matrices whose poles correspond to the masses of the scalar and axial-vector diquarks. 
The nucleon vertex function and mass, $M_N$, are obtained by solving the homogeneous 
Faddeev equation for a quark and a diquark \cite{cloet:2005}.
Because we need to solve this equation many times to obtain self-consistency, we
approximate the quark exchange kernel by a momentum 
independent form (static approximation). 
This necessitates the introduction of an additional parameter, $c$,
as explained in Ref.~\cite{Bentz:2001vc}.

%-------------------------------------------------------------------------------------------------
\begin{figure}[tbp]
\centering\includegraphics[height=8.5cm,angle=90]{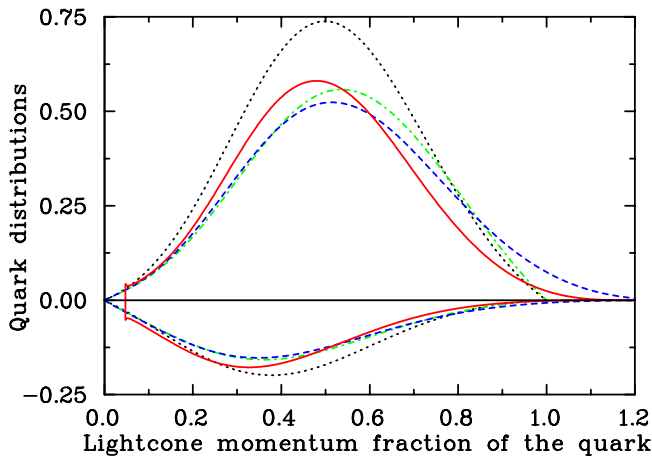}
\caption{Spin-dependent quark distributions, $\D u_v$ and $\D d_v$, 
at the model scale, $Q_0^2 = 0.16~\text{GeV}^2$.
There are four curves for each quark flavour, with the positive curves representing the
up distributions. The dotted line is the free nucleon distribution, the dot-dash line illustrates the 
effect of replacing the free masses with the effective ones. This distribution convoluted
with the Fermi smearing function, Eq.~(\ref{eqn:convolution}), is presented as the 
dashed line, and the final result where the vector field is also included via  
the scale transformation, Eq.~(\ref{eqn:vector}), is represented by the solid line.}
\label{fig:SD}
\vspace{-1.0em}
\end{figure}
%------------------------------------------------------------------------------------------------- 

To calculate the mean scalar and vector fields, we need the equation of state for nuclear matter. 
This can be rigorously derived for any
NJL Lagrangian using hadronization techniques, but in a simple mean-field approximation
the result for the energy density has the following form \cite{Bentz:2001vc}:
\begin{equation}
{\cal E} = {\cal E}_V - \frac{V_0^2}{4G_\omega} + 4\!\int\frac{d^3p}{(2\pi)^3}\,
\Theta\left(p_F - \vert\vec{p}\,\rvert\right)\varepsilon_p,
\end{equation}
where $\varepsilon_p = \sqrt{\vec{p}^{\,2} + M_N^2} + 3 V_0$ and the vacuum term ${\cal E}_V$
has the familiar ``Mexican hat'' shape.

%-------------------------------------------------------------------------------------------------
\begin{figure}[tbp]
\centering\includegraphics[height=8.5cm,angle=90]{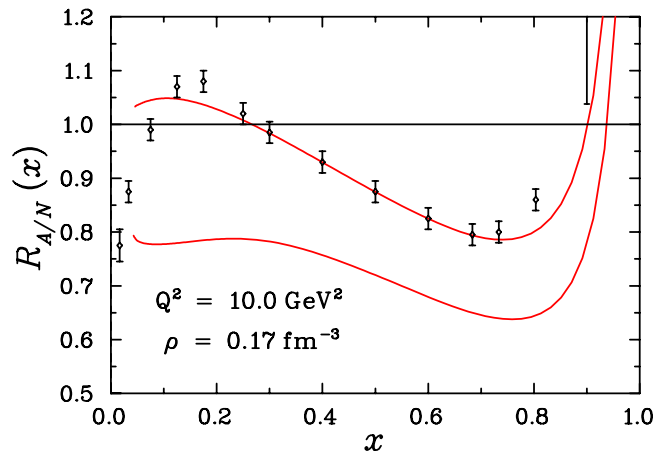}
\caption{Ratios of the spin-independent and spin-dependent nuclear to nucleon structure functions
at nuclear matter density.
The top curve is the usual EMC ratio $F_{2N}^A/F_{2N}$, where 
$F_{2N}$ is the isoscalar structure function and the superscript $A$ represents the in-medium result.
The EMC data for nuclear matter is taken from Ref.~\protect\cite{Sick:1992pw}.
Our prediction for the polarized EMC effect, $g_{1p}^A/g_{1p}$, is the lower
curve. Clearly we find a significant effect.}
\label{fig:EMC}
\end{figure}
%-------------------------------------------------------------------------------------------------

The parameters of the model are $\Lambda_{IR}$, $\Lambda_{UV}$, $M_0$, $c$, $G_\pi$, 
$G_s$, $G_a$ and $G_\omega$, where $\Lambda_{IR}$ and $\Lambda_{UV}$ are the 
infrared and ultraviolet cutoffs used in the proper-time regularization.
The infrared scale is expected to be of the order $\L_{QCD}$ and we set it to $\Lambda_{IR} = 0.28\,$GeV.
We also choose the free constituent quark mass to be $M_0 = 400\,$MeV \footnote{Our results 
do not depend strongly on this choice, remaining almost unchanged with $M_0$ is between 350 and 450 MeV.} and use
this constraint to fix the static parameter, $c$. The remaining
six parameters are fixed by requiring $f_\pi=93~$MeV, $m_\pi=140~$MeV,
$M_N = 940\,$MeV, the saturation point of nuclear matter 
$\lf(\rho_B,E_B\rg) = \lf(0.17\,\text{fm}^{-3},15.7\,\text{MeV}\rg)$
and lastly the Bjorken sum rule at zero density to be satisfied, with $g_A=1.267$.
We obtain $\Lambda_{UV} = 0.66\,$GeV, $c = 0.95\,$GeV, $G_\pi = 17.81\,
$GeV$^{-2}$, $G_s = 8.41\,$GeV$^{-2}$, $G_a = 1.36\,$GeV$^{-2}$ and  $G_\o = 5.58\,$GeV$^{-2}$.

With these model parameters the diquark masses at zero density are $M_s=0.65\,$GeV and $M_a=1.2\,$GeV
and vector field strength is $V_0 = 0.044\,$GeV. At saturation density
the effective masses become $M^* = 0.32\,$GeV, $M_s^* = 0.52\,$GeV, $M_a^* = 1.1\,$GeV and $M_N^* = 0.75\,$GeV. 

The results for the $u$ and $d$ spin-dependent quark distributions, at the model scale, are presented in 
Fig.~\ref{fig:SD}. There are four curves for each quark flavour, representing
the different stages leading to the full nuclear matter result.

Using these quark distributions we are able to construct the structure functions, $g_{1p}$
and $g_{1p}^A$, where the superscript $A$ represents a structure function in the nuclear 
medium. Analogous results for the spin-independent quark distributions \cite{cloet:2005}
allow us to determine the isoscalar structure functions $F_{2N}$ and $F_{2N}^A$, and hence
determine the EMC effect. Evolving \cite{Hirai:1997gb} these distributions
to a scale of 10$\,$GeV$^2$, we give in Fig.~\ref{fig:EMC} our results for the
ratios $F_{2N}^A/F_{2N}$ and $g_{1p}^A/g_{1p}$, that is the EMC and the polarized EMC effect. 
In the valence quark region, the model is able to reproduce the spin-independent EMC data 
extremely well. For the polarized ratio we find a significant effect, of the order
twice the size of the unpolarized EMC effect. 

%-------------------------------------------------------------------------------------------------
\begin{figure}[tbp]
\centering\includegraphics[height=8.5cm,angle=90]{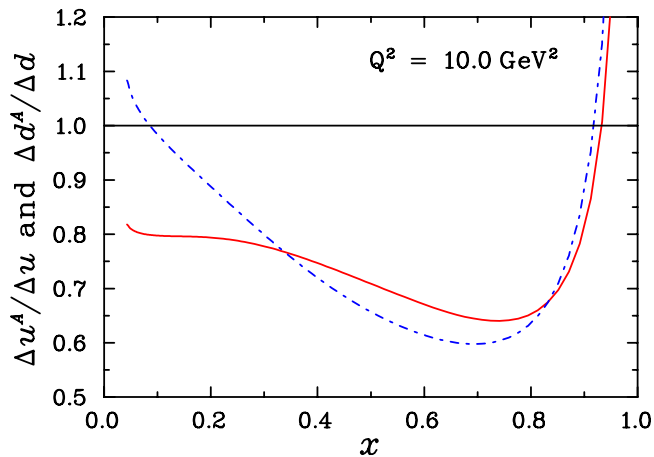}
\caption{Ratio of the quark distributions in nuclear matter to the
corresponding free distributions, at a scale of $Q^2 = 10\,$GeV$^2$. The solid line represents
$\D u^A(x)/\D u(x)$ and the dot-dashed line $\D d^A(x)/\D d(x)$. 
Note, these distributions are the full quark distributions and
hence include anti-quarks generated through $Q^2$ evolution.}
\label{fig:EMCflavour}
\end{figure}
%-------------------------------------------------------------------------------------------------

The nuclear quenching effects on the individual quark flavours is presented
in Fig.~\ref{fig:EMCflavour}. We find that the effect on both the $u$ and $d$ distributions
is large and approximately equal over the valence quark region. The resemblance between
$g_{1p}^A/g_{1p}$ and the ratio $\D u^A(x)/\D u(x)$ is simply because the up
distribution is enhanced by a factor 4 relative to the down and strange 
distributions in proton structure functions. Absent from our model
is the $U$(1) axial anomaly and sea quarks (at the model scale),
which prevents a reliable description of structure
functions at low $x$. For this reason in Figs. \ref{fig:EMC} 
and \ref{fig:EMCflavour} we do not plot our results in this region.

A thorough understanding of how nuclear medium effects arise from the
fundamental degrees of freedom $-$ the quarks and gluons $-$ represents 
an important challenge for the nuclear physics community.  
An experimental measurement of the polarized EMC effect would be another important
step toward this goal, providing important insights
into the quark polarization degrees of freedom within a nucleus. 
Our prediction of a remarkably large signature 
suggests that this  measurement is
feasible, and if these results are confirmed experimentally would yield
vital, new information on quark dynamics in the nuclear medium.

IC thanks W. Melnitchouk for helpful conversations.
This work was supported by the Australian Research Council and DOE contract DE-AC05-84ER40150,
under which SURA operates Jefferson Lab, and by the Grant in Aid for Scientific
Research of the Japanese Ministry of Education, Culture, Sports, Science and
Technology, Project No. C2-16540267.


\begin{thebibliography}{40}

%\cite{Aubert:1983xm}
\bibitem{Aubert:1983xm}
  J.~J.~Aubert {\it et al.}  [European Muon Collaboration],
  %``The Ratio Of The Nucleon Structure Functions F2 (N) For Iron And Deuterium,''
  Phys.\ Lett.\ B {\bf 123}, 275 (1983).
  %%CITATION = PHLTA,B123,275;%%

%\cite{Bodek:1983ec}
\bibitem{Bodek:1983ec}
  A.~Bodek {\it et al.},
  %``A Comparison Of The Deep Inelastic Structure Functions Of Deuterium And
  %Aluminum Nuclei,''
  Phys.\ Rev.\ Lett.\  {\bf 51}, 534 (1983).
  %%CITATION = PRLTA,51,534;%%

%\cite{Arnold:1983mw}
\bibitem{Arnold:1983mw}
  R.~G.~Arnold {\it et al.},
  %``Measurements Of The A-Dependence Of Deep Inelastic Electron Scattering From Nuclei,''
  Phys.\ Rev.\ Lett.\  {\bf 52}, 727 (1984).
  %%CITATION = PRLTA,52,727;%%

%\cite{Geesaman:1995yd}
\bibitem{Geesaman:1995yd}
  D.~F.~Geesaman, K.~Saito and A.~W.~Thomas,
  %``The nuclear EMC effect,''
  Ann.\ Rev.\ Nucl.\ Part.\ Sci.\  {\bf 45}, 337 (1995).
  %%CITATION = ARNUA,45,337;%%

%\cite{Smith:2002ci}
\bibitem{Smith:2002ci}
  J.~R.~Smith and G.~A.~Miller,
  %``Return of the EMC effect: Finite nuclei,''
  Phys.\ Rev.\ C {\bf 65}, 055206 (2002).
  %[arXiv:nucl-th/0202016].
  %%CITATION = NUCL-TH 0202016;%%

%\cite{Benesh:2003fk}
\bibitem{Benesh:2003fk}
  C.~J.~Benesh, T.~Goldman and G.~J.~Stephenson,
  %``Valence quark distribution in A = 3 nuclei,''
  Phys.\ Rev.\ C {\bf 68}, 045208 (2003).
  %[arXiv:nucl-th/0307038].
  %%CITATION = NUCL-TH 0307038;%%

%\cite{Guichon:2004xg}
\bibitem{Guichon:2004xg}
  P.~A.~M.~Guichon and A.~W.~Thomas,
  %``Quark structure and nuclear effective forces,''
  Phys.\ Rev.\ Lett.\  {\bf 93}, 132502 (2004).
  %[arXiv:nucl-th/0402064].
  %%CITATION = NUCL-TH 0402064;%%

%\cite{Ashman:1987hv}
\bibitem{Ashman:1987hv}
  J.~Ashman {\it et al.}  [European Muon Collaboration],
  %``A Measurement Of The Spin Asymmetry And Determination Of The Structure
  %Function G(1) In Deep Inelastic Muon Proton Scattering,''
  Phys.\ Lett.\ B {\bf 206}, 364 (1988).
  %%CITATION = PHLTA,B206,364;%%

%\cite{Chanfray:2004ev}
\bibitem{Chanfray:2004ev}
  G.~Chanfray and M.~Ericson,
  %``Nuclear matter saturation in a relativistic chiral theory and QCD
  %susceptibilities,''
  arXiv:nucl-th/0402018.
  %%CITATION = NUCL-TH 0402018;%%

%\cite{Bentz:2001vc}
\bibitem{Bentz:2001vc}
  W.~Bentz and A.~W.~Thomas,
  %``The stability of nuclear matter in the Nambu-Jona-Lasinio model,''
  Nucl.\ Phys.\ A {\bf 696}, 138 (2001).
  %[arXiv:nucl-th/0105022].
  %%CITATION = NUCL-TH 0105022;%%

%\cite{Mineo:2003vc}
\bibitem{Mineo:2003vc}
  H.~Mineo, W.~Bentz, N.~Ishii, A.~W.~Thomas and K.~Yazaki,
  %``Quark distributions in nuclear matter and the EMC effect,''
  Nucl.\ Phys.\ A {\bf 735}, 482 (2004).
  %[arXiv:nucl-th/0312097].
  %%CITATION = NUCL-TH 0312097;%%

%\cite{Schwinger:1951nm}
\bibitem{Schwinger:1951nm}
  J.~S.~Schwinger,
  %``On Gauge Invariance And Vacuum Polarization,''
  Phys.\ Rev.\  {\bf 82}, 664 (1951).
  %%CITATION = PHRVA,82,664;%%

%\cite{Ebert:1996vx}
\bibitem{Ebert:1996vx}
  D.~Ebert, T.~Feldmann and H.~Reinhardt,
  %``Extended NJL model for light and heavy mesons without q anti-q  thresholds,''
  Phys.\ Lett.\ B {\bf 388}, 154 (1996).
  %[arXiv:hep-ph/9608223].
  %%CITATION = HEP-PH 9608223;%%

%\cite{Hellstern:1997nv}
\bibitem{Hellstern:1997nv}
  G.~Hellstern, R.~Alkofer and H.~Reinhardt,
  %``Diquark confinement in an extended NJL model,''
  Nucl.\ Phys.\ A {\bf 625}, 697 (1997).
  %[arXiv:hep-ph/9706551].
  %%CITATION = HEP-PH 9706551;%%

%\cite{cloet:2005}
\bibitem{cloet:2005}
  I. C. Clo\"et, W. Bentz and A. W. Thomas,
  %``Quark structure and nuclear effective forces,''
  to appear.

%\cite{Saito:2001gv}
\bibitem{Saito:2001gv}
  K.~Saito, M.~Ueda, K.~Tsushima and A.~W.~Thomas,
  %``Structure Functions of Unstable Lithium Isotopes,''
  Nucl.\ Phys.\ A {\bf 705}, 119 (2002).
  %[arXiv:nucl-th/0110024].
  %%CITATION = NUCL-TH 0110024;%%

%\cite{Jaffe:1985je}
\bibitem{Jaffe:1985je}
  R.~L.~Jaffe,
  %``Deep Inelastic Scattering With Application To Nuclear Targets,''
  MIT-CTP-1261
  %\href{http://www.slac.stanford.edu/spires/find/hep/www?r=mit-ctp-1261}{SPIRES entry}
  {\it Lectures presented at the Los Alamos School on Quark Nuclear Physics, Los Alamos, N.Mex., Jun 10-14, 1985}.

%\cite{Barone:2001sp}
\bibitem{Barone:2001sp}
  V.~Barone, A.~Drago and P.~G.~Ratcliffe,
  %``Transverse polarisation of quarks in hadrons,''
  Phys.\ Rept.\  {\bf 359}, 1 (2002).
  %[arXiv:hep-ph/0104283].
  %%CITATION = HEP-PH 0104283;%%

%\cite{Ishii:1995bu}
\bibitem{Ishii:1995bu}
  N.~Ishii, W.~Bentz and K.~Yazaki,
  %``Baryons in the NJL model as solutions of the relativistic Faddeev
  %equation,''
  Nucl.\ Phys.\ A {\bf 587}, 617 (1995).
  %%CITATION = NUPHA,A587,617;%%

%\cite{Hirai:1997gb}
\bibitem{Hirai:1997gb}
  M.~Hirai, S.~Kumano and M.~Miyama,
  %``Numerical solution of Q**2 evolution equations for polarized structure functions,''
  Comput.\ Phys.\ Commun.\  {\bf 108}, 38 (1998).
  %[arXiv:hep-ph/9707220].
  %%CITATION = HEP-PH 9707220;%%

%\cite{Sick:1992pw}
\bibitem{Sick:1992pw}
  I.~Sick and D.~Day,
  %``The EMC effect of nuclear matter,''
  Phys.\ Lett.\ B {\bf 274}, 16 (1992).
  %%CITATION = PHLTA,B274,16;%%

\end{thebibliography}
\end{document}